\documentclass[pre,twocolumn,showpacs,amsmath,amssymb]{revtex4}

\usepackage{graphicx}
\usepackage{dcolumn}
\usepackage{bm}
\usepackage[usenames]{color}

\begin{document}

\preprint{APS/123-QED}

\title{Kinetic models of opinion formation in the presence \\ of personal conviction}

\author{Carlo Brugna}
\email{cabrugn@tim.it}
\author{Giuseppe Toscani}
\homepage{www-dimat.unipv.it/toscani} \email{giuseppe.toscani@unipv.it}
\affiliation{Dipartimento di Matematica,
  Universit\`a di Pavia,
  Via Ferrata 1,
  27100 Pavia, Italy.}

\date{\today}

\begin{abstract}
We consider a nonlinear kinetic equation of Boltzmann type which takes into account
the influence of conviction during the formation of opinion  in a system of agents
which interact through the binary exchanges introduced in   [G.\ Toscani,
\emph{Commun. Math. Sci.} \textbf{4},   481(2006)]. The original exchange mechanism,
which is based on the human tendency to compromise and change of opinion through
self-thinking, is here modified in the parameters of the compromise and diffusion
terms, which now are assumed to depend on the personal degree of conviction. The
numerical simulations show that the presence of conviction has the potential to break
symmetry, and to produce clusters of opinions. The model is partially inspired by the
recent work [L.\ Pareschi, G.\ Toscani, \emph{Phil. Trans. R. Soc. A} \textbf{372},
20130396 (2014)], in which the role of knowledge in the formation of wealth
distribution has been investigated.
\end{abstract}

\maketitle

\def\BigO{ {\mathcal {O}}}
\def\bigM{ {\mathcal {M}}}
\def\bigF{ {\mathcal {F}}}
\def\B{\hat \beta}
\def\f{\hat f}
\def\g{\hat g}
\def\Q{\hat Q}
\def\real{\mathbb{R}}
\newcommand{\R}{\mathbb R}
\newcommand{\N}{\mathbb N}
\newcommand{\dt}{\Delta t}
\def\be#1\ee{\begin{equation}#1\end{equation}}
\newcommand{\fer}[1]{(\ref{#1})}
\newtheorem{theorem}{Theorem}[section]
\newtheorem{lemma}[theorem]{Lemma}
\newtheorem{definition}[theorem]{Definition}
\newtheorem{assumption}{Assumption}
\newtheorem{nproposition}[theorem]{Proposition}
\newtheorem{remark}[theorem]{Remark}
\newtheorem{Co}[theorem]{Corollary}
\newtheorem{run}[theorem]{Run}
\newtheorem{example}[theorem]{Example}
\newtheorem{algorithm}{Algorithm}

\renewcommand{\theequation}{\arabic{section}.\arabic{equation}}
\renewcommand{\thetable}{\arabic{section}.\arabic{table}}
\renewcommand{\thefigure}{\arabic{section}.\arabic{figure}}
\newcommand{\bq}{\begin{equation}}
\newcommand{\eq}{\end{equation}}


\def\nnew{\color{red}}

\def\reall{\mathfrak{I}}
\def\bqa{\begin{eqnarray}}
\def\eqa{\end{eqnarray}}
\def\dt{\Delta t}
\def\m{\mu}
\def\r{\rho}
\def\l{\lambda}
\def\g{\gamma}
\def\e{\epsilon}
\def\a{\alpha}
\def\d{\delta}
\def\t{\theta}
\def\b{\beta}
\def\S{\sigma}
\def\r{\rho}


\newcommand{\dis}{\displaystyle}
\newcommand{\bd}{\begin{displaymath}}
\newcommand{\ed}{\end{displaymath}}
\newcommand{\ba}{\begin{eqnarray}}
\newcommand{\ea}{\end{eqnarray}}
\newcommand{\doubleint}{\int\!\!\!\!\int}
\newcommand{\tripleint}{\int\!\!\!\!\int\!\!\!\!\int}
\newcommand{\s}{\sum_{s\,=1}^{4}}
\newcommand{\er}{\sum_{r\,=1}^{4}}
\newcommand{\sr}{\sum_{s,r\,=1}^{4}}
\newcommand{\p}{\partial}
\newcommand{\effe}{{\bf f_{\atop{\,\,}{\!\!\!\!\!\!\!{\tilde{}}}}}}


\def\ff{\widetilde f}
\def\lf{{\mathcal L}f}
\def\bb{\hat \beta}
\def\MM{\hat M}
\def\gg{\hat g}
\def\Realr{\mathbb{R}}
\def\N{\mathbb{N}}
\def\R{\mathbb{R}}
\def\var{\varepsilon}
\def\pa{\partial}
\newcommand{\setR}{\mathbb{R}}
\newcommand{\theq}{{\mathbf S}}
\newcommand{\ther}{{\mathbf Q}}
\def\gl{ \widetilde{g}}
\def\Ci{P(|v|)}
\def\Cii{P(|w|)}
\def\Di{D(|v|)}
\def\Dii{D(|w|)}
\def\Ball{ {\mathcal {B}}}
\def\Chi{\chi}
\newenvironment{equations}{\equation\aligned}{\endaligned\endequation}

\section{Introduction}

In recent years, the dynamics of opinion formation in a multi-agent society  has
received growing attention \cite{BN2, BN3, BN1, Def, GZ,GGS, SW}. In reason of its
cooperative nature, it appeared natural to resort to tools and methods typical of
statistical mechanics to study such systems \cite{CFL,Cha}. The approaches considered
so far range from cellular automata, especially used for numerical simulation, to
models of mean field type, which lead to systems of (ordinary or partial) differential
equations, to kinetic models of opinion formation \cite{BeDe, Bou, Bou1, Bou2, CDT,
To06}. In kinetic models, the variation of opinion is obtained through binary
interactions between agents. In view of the relation between parameters in the
microscopic binary rules, the society develops a certain steady macroscopic opinion
distribution \cite{MT, PT13}, which characterizes the formation of a relative
consensus around certain opinions.

The relevant aspects to be taken into account when modelling  binary interactions in
opinion formation have been identified in the compromise process \cite{Def, Wei}, in
which individuals tend to reach a compromise after exchange of opinions, and the
self-thinking \cite{To06}, where individuals change their opinion in a diffusive
fashion.

Following this line of thought,  a wide class of kinetic models of opinion formation,
based on two-body interactions involving both compromise and diffusion properties in
exchanges between individuals, has been introduced in~\cite{To06}.  These models are
sufficiently general to take into account a large variety of human behaviors, and to
reproduce in many cases explicit steady profiles from which one can easily elaborate
information on the macroscopic opinion distribution. This type of modelling has
subsequently been applied to various situations in \cite{Bou,Bou1, Bou2, DMPW}.

More recently, a further relevant parameter, strongly related to the problem of
opinion formation,  has been taken into account in \cite{LCCC}. Resembling the model
for wealth exchange in a multi-agent society introduced in \cite{ChaCha00}, this new
has an additional  parameter to quantify the personal \emph{conviction}, representing
a measure of the influencing ability of individuals. Individuals with high conviction
are resistant to change opinion, and have a prominent role in attracting other
individuals towards their opinions. In this sense, individuals with high conviction
play the role of leaders \cite{DMPW}.

The role of conviction has been subsequently considered by other authors. It was shown
in \cite{LCCC} that beyond a certain value of this conviction parameter, the society
reaches a consensus, where one of the two choices (positive or negative) provided to
the individuals prevails, thereby spontaneously breaking a discrete symmetry. A
further model in which this parameter has been taken into account was proposed in
\cite{Sen}. There, the self-conviction and the ability to influence others were taken
as independent variables.  Also, exact solutions of a discrete opinion formation model
with conviction were found by Biswas in \cite{Bis}. In \cite{Cro1, Cro}, conviction
has been introduced as relevant parameter in  a class of discrete opinion models.
Within this class, each agent opinion takes only discrete values,  and its time
evolution is ruled by two terms, one representing binary interactions between
individuals, and the other the degree of conviction or persuasion (a
self-interaction).

In all the aforementioned models, conviction is realized by a fixed-in-time parameter
(or a random variable), eventually different for different individuals. Consequently,
while it is clear that a certain distribution of this parameter among agents leads to
a steady distribution of opinions with properties  which are related to it, it is not
completely clarified why conviction has to be assumed with a certain distribution.

All these studies, however, indicate that, among the various behavioral aspects that
determine a certain opinion formation, conviction represents an important variable to
be taken into account. This problem has deep analogies with a recent study of one of
the present authors with Pareschi \cite{PaTo14}, where the role of knowledge in wealth
distribution forming has been studied by allowing agents to depend on two variables,
denoting knowledge and wealth respectively. There, a kinetic equation for the
evolution of knowledge has been coupled with a kinetic model for wealth distribution,
by allowing binary trades to be dependent of the personal degree of knowledge.
\emph{Mutatis mutandis}, we will assume in this paper that individuals are
characterized by two variables, representing conviction and, respectively, opinion.
Following the line of thought in \cite{PaTo14}, we will introduce here a kinetic model
for conviction formation, by assuming that the way in which conviction is formed is
independent of the personal opinion. Then, the (personal) conviction parameter will
enter into the microscopic binary interactions for opinion formation considered in
\cite{To06}, to modify them in the compromise and self-thinking terms. Within this
picture, both the conviction and the opinion are modified in time in terms of
microscopic interactions. As we shall see by numerical investigation, the role of the
additional conviction variable is to bring the system towards a steady distribution in
which there is formation of clusters even in absence of bounded confidence hypotheses
\cite{BN2,BN3, BN1, HK}.

In our kinetic model for conviction formation, we will assume that the relevant terms
responsible of the modification of the conviction are from one side the the
acquisition of information, and from the other side the possibility of afterthought
and to think back, which appear to be natural and universal features. In reason of
this, the positive parameter that quantifies conviction can either increase (through
information) or decrease (through afterthought).  The relevant aspect is that
information will be achieved though the surrounding, thus producing a linear kinetic
model. Note that it is assumed here that the individual conviction is not correlated
to personal opinion, thus allowing formation of conviction without resorting to the
distribution of opinions. Then, the linear conviction interaction will be coupled with
the binary exchange of opinion introduced in \cite{To06}, which includes both the
compromise propensity and the the change of the personal opinion due to self-thinking.
In the new interaction rule both the compromise and the self-thinking part of the
opinion exchange depend on the personal conviction. A typical and natural assumption
is that high conviction could act on the interaction process both to reduce the
personal propensity to compromise, and to reduce the self-thinking in the interaction.
These rules will subsequently be merged, within the principles of classical kinetic
theory, to derive a nonlinear Boltzmann-like kinetic equation for the joint evolution
of conviction and opinion variables.

The kinetic approach revealed to be a powerful tool \cite{NPT, PT13}, complementary to
the numerous theoretical and numerical studies that can be found in the recent
physical and economic literature on these subjects. On the other hand, as many other
approaches, the study of the socio-economic behavior of a (real) population of agents
by means of kinetic models with very few (essential) parameters is able to capture
only partially the extremely complex behavior of such systems. The idea to introduce
as additional parameter the conviction  in the study of opinion formation goes exactly
into the direction to give a more accurate description of the human behavior. Not
surprisingly, the description of the evolution of pair conviction--opinion in terms of
a kinetic equation gives rise to a variety of challenging mathematical problems, both
from the theoretical and numerical point of view. In particular, concerning the
distribution of conviction, it is remarkable that this class of simple models is able
to reproduce various features always present in the reality, like the presence of a
considerable number of undecided in the population, as well as the formation of
clusters of opinions in the steady distribution.

To end this introduction, we outline that the problem of clustering  is of
pa\-ra\-mount importance in this context. Indeed, models for opinion formation belong
to the variety of  models for self-organized dynamics in social, biological, and
physical sciences \cite{NPT}, which assume that the intensity of alignment increases
as agents get closer, reflecting a common tendency to align with those who think or
act alike. As noticed by Motsch and Tadmor \cite{MT} \emph{similarity breeds
connection} reflects our intuition that increasing the intensity of alignment as the
difference of positions decreases is more likely to lead to a consensus. However, it
is argued in \cite{MT}  that the converse is true: when the dynamics is driven by
local interactions, it is more likely to approach a consensus when the interactions
among agents increase as a function of their difference in position. In absence of
further parameters, heterophily, the tendency to bond more with those who are
different rather than with those who are similar, plays a decisive role in the process
of clustering. This motivates further our choice to resort to the additional role of
conviction.

The paper is organized as follows. In Section \ref{know6} we introduce
and discuss the linear kinetic model for the formation of conviction in a multi-agent society. This linear model is based on microscopic interactions with a fixed background, and is such that the density of the population conviction converges towards a steady distribution which is heavily dependent of the microscopic parameters of the microscopic interactions. Then, the conviction rule is merged with the binary  interaction for opinion to obtain a nonlinear kinetic model of Boltzmann-type for the joint density of conviction and opinion. This part is presented in Section  \ref{model}.  Last, Section \ref{nume} is devoted to various numerical experiments, which allow to recover the steady joint distribution of conviction and opinion in the population for various choices of the relevant parameters.

\section{The formation of conviction}\label{know6}

To give a precise and well-established definition of conviction  is beyond our
purposes. Instead of resorting to a definition it seems natural to agree on certain
universal aspects about it. Conviction can be described as a certain resistance to
modify a personal behavior. How the personal amount of conviction is formed is a very
difficult question. We can reasonably argue that, among other reasons, responsible of
conviction forming include familiar environment, personal contacts, readings or skills
acquired through experience or education. It is natural to assume that conviction (at
least concerning some aspects of life like religious or political beliefs) is in part
inherited in the interior of family from the parents, but it is also evident that the
the main factor that can influence it is the social background in which the individual
grows and lives \cite{TB}. Indeed, the experiences that lead to be convinced about
something can not be fully inherited from the parents, such as the eye color, but
rather are acquired by several elements of the environment. This process is manifold
and produces different results for each individual in a population. Like in knowledge
formation, although all individuals are given the same opportunities, at the end of
the process every individual appears to have a different level of conviction about
something. Also, it is almost evident that the personal conviction is heavily
dependent on the individual nature. A consistent part of us is accustomed to rethink,
and to have continuous afterthoughts on many aspects of our daily decisions. This is
particularly true nowadays, where the global access to information via web gives to
each individual the possibility to have a \emph{reservoir} of infinite capacity from
which to extract any type of (useful or not) information, very often producing
insecurity.

The previous remarks are at the basis of a suitable description of the evolution of
the distribution of conviction in a population of agents by means of microscopic
interactions with a fixed background. We will proceed as in \cite{PaTo14}. Each
variation of conviction is interpreted as an interaction where a fraction of the
conviction of the individual is lost by virtue of afterthoughts and insecurities,
while at the same time the individual can absorb  a certain amount of conviction
through the information achieved from the external background (the surrounding
environment). In our approach, we quantify the conviction of the individual in terms
of a scalar parameter $x$, ranging from zero to infinity.  Denoting with $z \ge 0$ the
degree of conviction achieved from the background, it is assumed that the new amount
of conviction  in a single interaction can be computed as
 \be\label{k1}
 x^* = (1-\lambda(x))x + \lambda_B(x) z + \kappa H(x).
 \ee
In \fer{k1} the functions $\lambda= \lambda(x)$ and $\lambda_B= \lambda_B(x)$
quantify, respectively, the personal amounts of insecurity and willingness to be
convinced by others, while $\kappa$ is a random parameter which takes into account the
possible unpredictable modifications of the conviction process. We will in general fix
the mean value of $\kappa$ equal to zero. Last, $H(\cdot)$ will denote an increasing
function of conviction. The typical choice is to take $H(x)= x^\nu$, with $0<\nu\le
1$. Since some insecurity is always present, and at the same time it can not exceed a
certain amount of the total conviction, it is assumed that $\lambda_- \le \lambda(x)
\le \lambda_+$, where $\lambda_- >0$, and $\lambda_+ < 1$. Likewise, we will assume an
upper bound for the willingness to be convinced by the environment. Then, $0 \le
\lambda_B(x)\le \bar\lambda$, where $\bar\lambda <1$.  Lastly, the random part is
chosen to satisfy the lower bound $\kappa \ge - (1-\lambda_+)$. By these assumptions,
it is assured that the post-interaction value
 $x^*$ of the conviction is nonnegative.

Let $C(z)$, $z \ge 0$ denote the probability distribution of
degree of conviction of the (fixed) background. We will suppose that
$C(z)$ has a bounded mean, so that
 \be\label{ba1}
 \int_{\R_+} C(z) \, dz = 1; \quad \int_{\R_+} z\,C(z) \, dz = M
  \ee
We note that the distribution of the background will induce a certain policy of
acquisition of conviction. This aspect has been discussed in \cite{PaTo14}, from which
we extract the example that follow. Let us assume that the background is a random
variable uniformly distributed on the interval $(0, a)$, where $a > 0$ is a fixed
constant. If we choose for simplicity $\lambda(x) = \lambda_B(x)= \bar\lambda$, and
the individual has a degree of conviction $x>a$, in absence of randomness the
interaction  will always produce a value $x^* \le x$, namely a partial decrease of
conviction. In this case, in fact, the process of insecurity in an individual with
high conviction can not be restored by interaction with the environment.

The study of the
time-evolution of the distribution of conviction produced by binary
interactions  of type \fer{k1}  can be obtained by
resorting to kinetic collision-like models \cite{PT13}. Let $F=
F(x,t)$ the density of agents which at time $t >0$ are represented by their
conviction $x \in \R_+$. Then, the time
evolution of $F(x, t)$  obeys to a
Boltzmann-like equation. This equation is usually written
in weak form. It corresponds to say that the solution $F(x,t)$
satisfies, for all smooth functions $\varphi(x)$ (the observable quantities)
 \begin{equations}
  \label{kine-w}
 &\frac{d}{dt}\int_{\R_+}F(x,t)\varphi(x)\,dx  = \\
  & \Big \langle \int_{\R_+^2} \bigl( \varphi(x^*)-\varphi(x) \bigr) F(x,t)C(z)
\,dx\,dz \Big \rangle.
 \end{equations}
In \fer{kine-w}  the post-interaction conviction $x^*$ is given by  \fer{k1}. As
usual, $\langle \cdot \rangle$ represents mathematical expectation. Here expectation takes into account the presence of the random parameter $\kappa$ in \fer{k1}.

The meaning of the kinetic equation \fer{kine-w} is the following. At any positive time $t >0$, the variation in time of the distribution of conviction (the left-hand side) results from a balance equation in which, through interaction with the background we gain agents with conviction $x^*$  loosing agents with conviction $x$.  This change is measured by the interaction operator at the right-hand side.

In order to verify if the kinetic equation \fer{kine-w} gives reasonable outputs on the distribution of conviction among the population of agents, let us study some of its properties.
It is immediate to recognize that equation \fer{kine-w} preserves the
total mass, so that $F(x,t)$, $t >0$, remains a probability density if it is so
initially. By choosing $\varphi(x) = x$ we recover the evolution of
the mean conviction $M_C(t)$ of the agents system, which gives a first measure of the conviction rate. The mean value satisfies the equation
 \be\label{m2}
 \frac{dM_C(t)}{dt} = - \int_{\R_+} x \lambda(x) F(t) \, dx +
 M \int_{\R_+}  \lambda_B(x) F(t) \, dx,
 \ee
which in general it is not explicitly solvable, unless the functions $\lambda(x)$ and
$\lambda_B(x)$ are assumed to be constant. However, since $\lambda(x) \ge \lambda_-$,
while $ \lambda_B(x) \le \bar\lambda$, the mean value always satisfies the
differential inequality
 \be\label{m3}
 \frac{dM_C(t)}{dt} \le - \lambda_- M_C(t) + \bar\lambda M,
 \ee
which guarantees that the mean conviction of the system will never
exceed the (finite) value $M_{max}$ given by
 \[
M_{max} = \frac{ \bar\lambda}{\lambda_-} M.
 \]
If  $\lambda(x)= \lambda $ and $\lambda_B(x)=
\lambda_B$ are constant, equation \fer{m2}
becomes
 \be\label{m5}
 \frac{dM_C(t)}{dt} = - \lambda M_C(t) +
  \lambda_B M.
 \ee
In this case, the linear differential equation can be solved, and
 \be\label{ex1}
M_C(t) = M_C(0) e^{-\lambda t} + \frac{\lambda_B M}{\lambda} \left(
 1- e^{-\lambda t} \right).
 \ee
Formula \fer{ex1} shows that the mean conviction converges exponentially to its limit
value $\lambda_B M/\lambda$. Note that by increasing the parameter $\lambda$ which
measures  the personal amount of insecurity  we decrease the final mean conviction.

A further insight into the linear kinetic equation \fer{kine-w} can be obtained by
resorting to particular asymptotics which lead to Fokker-Planck equations
\cite{CoPaTo05}. In order to describe the asymptotic process, let us discuss in some
details the evolution equation for the mean conviction, given by \fer{m5}. For
simplicity, and without loss of generality, let us assume $\lambda$ and $ \lambda_B$
constant. Given a small parameter $\e$, the scaling
 \be\label{scal}
\lambda \to \e\lambda,\quad \lambda_B \to \e \lambda_B, \quad \kappa
\to \sqrt\e \kappa
 \ee
 is such that the mean value $M_C(t)$ satisfies
 \[
\frac{dM_C(t)}{dt} = -\e\left( \lambda M_C(t) -
  \lambda_B M\right).
 \]
If we set $\tau = \e t$, $F_\e(x,\tau) = F(x,t)$, then
 \[
M_C(\tau) = \int_{\R_+} x F_\e(x,\tau) \, dx = \int_{\R_+} x
F(x,\tau) \, dx = M_C(t),
 \]
and the mean value of the density $F_\e(x,\tau)$ solves
  \be\label{meanscal}
 \frac{dM_C(\tau)}{d\tau} = -  \lambda M_C(\tau) +
  \lambda_B M.
 \ee
Note that equation \fer{meanscal} does not depend explicitly on the
scaling parameter $\e$. In other words, we can reduce in each
interaction the variation of conviction, waiting enough time to get
the same law for the mean value of the knowledge density.

We can consequently investigate the situation in which most of the
interactions produce a very small variation of conviction ($\e \to
0$), while at the same time  the evolution of the conviction density
is such that \fer{meanscal} remains unchanged. We will call this
limit quasi-invariant conviction limit.

Let now assume that the centered random variable $\kappa$ has bounded moments at least
of order $n=3$, with $\langle \kappa^2 \rangle = \mu$. Then, proceeding as in
\cite{CoPaTo05} (cf. also Chapter 1 in \cite{PT13}), we obtain that the density
$F_\e(x,\tau)$ solves the equation
 \begin{equations}\label{FPe}
  &\frac{d}{dt}\int_{\R_+}F_\e(x,t)\varphi(x)\,dx =\\
  & - \int_{\R_+}( \lambda(x) x - \lambda_B(x) M ) + F_\e(x,t)\varphi'(x)\,dx \\
 & + \frac \mu 2 \int_{\R_+} H^2(x)F_\e(x,t)\varphi''(x)\,dx + R_\e(x,\tau),
 \end{equations}
where the remainder $R_\e$ is vanishing as $\e \to 0$. Consequently, it is shown that, as $\e \to 0$, the density $F_\e(x,\tau)$ converges towards the density $G(x, \tau)$ solution of the Fokker-Planck equation
 \begin{equations}\label{FP5}
 &\frac{\partial G(x,\tau)}{\partial \tau} = \frac \mu 2 \frac{\partial^2 }{\partial x}
 \left(H(x)^2 G(x,\tau)\right) + \\
 &\frac{\partial}{\partial x}\left( (\lambda(x) x - \lambda_B(x) M) G(x,\tau)\right).
 \end{equations}

The case in which $\lambda(x) = \lambda$ and $\lambda_B(x) = \lambda_B$ allows to get
the explicit form of the steady distribution of conviction \cite{PT13}. We will
present two realizations of the asymptotic profile, that enlighten the consequences of
the choice of a particular function $H(\cdot)$. First, let us consider the case in
which $H(x)=x$. In this case, the Fokker--Planck equation \fer{FP5} coincides with the
one obtained in \cite{CoPaTo05}, related to the steady distribution of wealth in a
multi--agent market economy. One obtains
 \be\label{stead1}
 G_\infty(x) = \frac{G_0}{x^{2 + 2\lambda/\mu}} \exp \left\{ - \frac{2\lambda_B M}{\mu x}\right\}.
 \ee
In \fer{stead1} the constant $G_0$ is chosen to fix the total mass of $G_\infty(x)$
equal to one. Note that the steady profile is heavy tailed, and the size of the
polynomial tails is related to both $\lambda$ and $\sigma$. Hence, the percentage of
individuals with high conviction is decreasing as soon as the parameter $\lambda$ of
insecurity is increasing, and/or the parameter of self-thinking is decreasing. It is
moreover interesting to note that the size of the parameter $\lambda_B$ is important
only in the first part of the $x$-axis, and contributes to determine the size of the
number of undecided. Like in the case of wealth distribution, this solution has a
large \emph{middle class}, namely a large part of the population with a certain degree
of conviction, and a small \emph{poor class}, namely a small part of undecided people.

The second case refers to the choice $H(x) = \sqrt x$. Now, people with high
conviction is more resistant to change (randomly) with respect to the previous case.
On the other hand, if the conviction is small, $x < 1$, the individual is less
resistant to change. Direct computations now show that the steady profile is given by
 \be\label{stead2}
 H_\infty(x) = {H_0}\,{x^{-1 + (2\lambda_B M)/\mu}} \exp \left\{ - \frac{2\lambda}{\mu} x \right\},
 \ee
where the constant $H_0$ is chosen to fix the total mass of $H_\infty(x)$ equal to
one. At difference with the previous case, the distribution decays exponentially to
infinity, thus describing a population in which there are very few agents with a large
conviction. Moreover, this distribution describes a population with a huge number of
undecided agents. Note that, since the exponent of $x$ in $H_\infty(\cdot)$ is
strictly bigger than $-1$, $H_\infty(\cdot)$ is integrable for any choice of the
relevant parameters.

Other choices of the exponent $\nu$ in the range $0<\nu\le 1$ do not lead to essential
differences.  The previous examples show that, despite the simplicity of the kinetic
interaction \fer{k1},  by acting on the coefficient of the random part $\kappa$ one
can obtain very different types of steady conviction distributions.

\section{The Boltzmann equation for opinion and conviction}\label{model}
\setcounter{equation}{0}

In this section, we will join the kinetic model for conviction with the kinetic model
for opinion formation introduced in  \cite{To06}. This model belongs to a class of
models in which agents are indistinguishable. In most of these models \cite{PT13} an
agent's \emph{state} at any instant of time $t\geq0$ is completely characterized by
his opinion $v$.
 In agreement with the usual assumptions of the
pertinent literature, the variable $v$  varies continuously from $-1$ to $1$, where
$-1$ and $1$ denote  two (extreme) opposite opinions. A remarkable consequence of
introducing a continuous distribution of opinions in the interval $[-1, 1]$ is that
the extremal opinions do not play any particular rule.

The unknown in this model is the density (or distribution function) $f = f(v, t)$,
where $v \in \reall = [-1, 1]$ and the time $t \ge 0$, whose time evolution is
described, as shown later, by a kinetic equation of Boltzmann type.

The precise meaning of the density $f$ is the following. Given the population to
study, if the opinions are defined on a sub-domain $D \subset \reall$ , the integral
 \[
 \int_D f(v,t)\, dv
 \]
represents the \emph{number} of individuals with opinion included in $D$ at time $t>
0$. It is assumed that the density function is normalized to $1$, that is
 \[
 \int_\reall f(v,t) \, dv = 1.
 \]
As always happens when dealing with a kinetic problem in which the variable belongs to
a bounded domain, this choice introduces supplementary mathematical difficulties in
the correct definition of binary interactions. In fact, it is essential to consider
only interactions that do not produce opinions outside the allowed interval, which
corresponds to imposing that the extreme opinions cannot be crossed. This crucial
limitation emphasizes the difference between the present {\em social} interactions,
where not all outcomes are permitted, and the classical interactions between
molecules, or, more generally,  the wealth trades (cf. \cite{PT13}, Chapter 5), where
the only limitation for trades was to insure that the post-collision wealths had to be
non-negative.

In order to build a possibly realistic model, this severe limitation has to be coupled
with a reasonable physical interpretation of the process of opinion forming. In other
words, the impossibility of crossing the boundaries has to be a by-product of  good
modelling of binary interactions.

From a microscopic viewpoint, the binary interaction in \cite{To06} are described by
the rules
\begin{equations}\label{ch6:trade_rule}
& v^* =  v - \gamma\Ci( v- w) + \Theta \Di, \\
&\null \\[-.25cm]
 & w^*   =  w - \gamma \Cii(w- v) + \tilde\Theta \Dii .
\end{equations}

In \fer{ch6:trade_rule}, the pair $(v,w)$, with $v, w \in \reall$, denotes the
opinions of two arbitrary individuals before the interaction, and $(v^*,w^*)$ their
opinions after exchanging information between each other and with the exterior. The
coefficient $\gamma\in (0,1/2)$ is a given constant, while $\Theta$ and $\tilde\Theta$
are random variables with the same distribution, with zero mean and variance
$\sigma^2$, taking values on a set $\Ball \subseteq \R$.  The constant $\gamma$ and
the variance $\sigma^2$ measure respectively the compromise propensity and the degree
of spreading of opinion due to diffusion, which describes possible changes of opinion
due to personal access to information (self-thinking). Finally, the functions
$P(\cdot)$ and $D(\cdot)$ take into account the local relevance of compromise and
diffusion for a given opinion.

Let us describe in detail the interaction on the right-hand side of
\fer{ch6:trade_rule}. The first part is related to the compromise  propensity of the
agents,  and the last contains the diffusion effects due to individual deviations from
the average behavior.  The presence of both the functions $P(\cdot)$ and $D(\cdot)$ is
linked to the hypothesis that openness to change of opinion is linked to the opinion
itself, and decreases as one gets closer to extremal opinions. This corresponds to the
natural idea that extreme opinions are more difficult to change. Various realizations
of these functions can be found in \cite{To06}. In all cases, however, we assume that
both $\Ci$ and $\Di$ are non-increasing with respect to $|v|$, and in addition $0 \le
\Ci \le 1$, $0 \le \Di \le 1$. Typical examples are given by $P(|v|) = 1-|v|$ and $D(|v|) =\sqrt{1-v^2}$.

In the absence of the diffusion contribution ($\Theta, \tilde\Theta\equiv 0$),
\fer{ch6:trade_rule} implies
  \begin{equations}\label{ch6:tr+-}
 & v^*+w^* = v+w  + \gamma (v-w) \left( \Ci - \Cii \right),\\
\\[-.25cm]
 & v^*-w^* = \left(1- \gamma( \Ci + \Cii)\right)(v-w).
 \end{equations}
Thus, unless the function $P(\cdot)$ is assumed constant, $P=1$, the total {\em
momentum} is not conserved and it can increase or decrease depending on the opinions
before the interaction. If $P(\cdot)$ is assumed constant, the conservation law is
reminiscent of analogous conservations which take place in kinetic theory. In such a
situation, thanks to the upper bound on the coefficient $\gamma$, equations
(\ref{ch6:trade_rule}) correspond to a granular-gas-like interaction \cite{PT13} where
the stationary state is a Dirac delta centered on the average opinion. This behavior
is a consequence of the fact that, in a single interaction, the compromise propensity
implies that the difference of opinion is diminishing, with $|v^*-w^*| =
(1-2\gamma)|v-w|$. Thus, all agents  in the society will end up with exactly the same
opinion. Note that in this elementary case a constant part of the relative opinion is
restituted after the interaction. In all cases, however, the second inequality in
\fer{ch6:tr+-} implies that the difference of opinion is diminishing after the
interaction.

We remark, moreover, that, in the absence of diffusion, the lateral bounds are not
violated, since
 \begin{equations}\label{ch6:gran-rule}
v^* & =  & (1 - \gamma\Ci)v + \gamma\Ci w, \\
\\[-.25cm]
 w^*  & =&   (1 - \gamma\Cii ) w + \gamma\Cii v,
 \end{equations}
 imply
 \[
 \max\left\{|v^*|, |w^*|\right\} \le  \max\left\{|v|,
 |w|\right\}.
 \]

Let  $f(v,t)$ denote the distribution of opinion $v \in \reall$ at time $t \ge 0$. The
time evolution of $f$ is recovered as a balance between bilinear gain and loss of
opinion terms, described in weak form by the integro-differential equation of
Boltzmann type
\begin{equations}\label{ch6:weak boltz}
&\frac d{dt}\int_{\reall} \varphi(v)f(v,t)\,dv  = (Q(f,f),\varphi) =  \\
 &\left\langle \int_{\reall^2}   f(v)
f(w) ( \varphi(v^*)+ \varphi(w^*)-\varphi(v)-\varphi(w)) d v d w \right\rangle,
\end{equations}
where $(v^*,w^*)$ are the post-interaction opinions generated by the pair $(v,w)$ in
\fer{ch6:trade_rule}.

We remark that equation \fer{ch6:weak boltz} is consistent with the fact that a
suitable choice of the function $D(\cdot)$ in \fer{ch6:trade_rule} coupled with a
small support $\Ball$ of the random variables implies that both $|v^*| \le 1$ and
$|w^*| \le 1$.

The analysis in \cite{To06} essentially shows that the microscopic interaction
\fer{ch6:trade_rule} is so general that the kinetic equation \fer{ch6:weak boltz} can
describe a variety of different behaviors of opinion. In its original formulation,
both the compromise  and the self-thinking intensities  were assigned in terms of the
universal constant $\gamma$ and of the universal random parameters
$\Theta,\tilde\Theta$. Suppose now that these quantities in \fer{ch6:trade_rule} could
depend of the personal conviction of the agent. For example, one reasonable assumption
would be that an individual with high personal conviction is more resistant to move
towards opinion of any other agent by compromise. Also, an high conviction could imply
a reduction of the personal self-thinking. If one agrees with these assumptions, the
binary trade \fer{ch6:trade_rule} has to be modified to include the effect of
conviction. Given two agents $A$ and $B$ characterized by the pair $(x,v)$
(respectively $(y,w)$) of conviction and opinion, the new binary trade between $A$ and
$B$ now reads
\begin{equations}
  \label{eq.cpt2}
& v^* =  v - \gamma\,\Psi(x)\Ci( v- w) + \Phi(x)\Theta \Di, \\
&\null \\[-.25cm]
 & w^*   =  w - \gamma\,\Psi(y) \Cii(w- v) + \Phi(y)\tilde\Theta \Dii .
\end{equations}
In \fer{eq.cpt2} the personal compromise propensity and self-thinking of the agents
are modified by means of the functions $\Psi =\Psi(x)$ and $\Phi= \Phi(x)$, which
depend on the convictions parameters. In this way, the outcome of the interaction
results from a combined effect of (personal) compromise propensity, conviction and
opinion. Among other possibilities, one reasonable choice is to fix the functions
$\Psi(\cdot)$ and $\Phi(\cdot)$ as non-increasing functions. This reflects the idea
that the conviction acts to increase the tendency to remain of the same opinion. Among others,
a possible choice is
 \[
\Psi(x) = (1+(x-A)_+)^{-\alpha}, \quad \Phi(x) = (1+(x-B)_+)^{-\beta}.
 \]
Here  $A, B, \alpha, \beta$ are nonnegative constants, and $h(x)_+$ denotes the positive part of $h(x)$. By choosing $A>0$ (respectively $B>0$), conviction will start to influence the change of opinion only when $x >A$ (respectively $x>B$).
It is interesting to remark that the presence of the conviction parameter (through the
functions $\Psi$ and $\Phi$), is such that the post-interaction opinion of an agent
with high conviction remains close to the pre-interaction opinion. This induces a
mechanism in which the opinions of agents with low conviction are attracted towards
opinions of agents with high conviction.

Assuming the binary trade \fer{eq.cpt2} as the microscopic binary exchange of
conviction and opinion in the system of agents, the joint evolution of these
quantities is described in terms of the density $f= f(x,v,t)$ of agents which at time
$t >0$ are represented by their conviction $x \in \R_+$ and wealth $v\in \reall$. The
evolution in time of the density $f$ is described by the following kinetic equation
(in weak form) \cite{PT13}
 \begin{equations}  \label{kine-xv}
&\frac{d}{dt}\int_{\R_+\times \reall}\varphi(x,v) f(x,v,t)\,dx\, dv  = \\
&\frac 12 \Big \langle \int_{\R_+^2\times \reall^2} \bigl( \varphi(x^*,v^*) +
\varphi(y^*,w^*) -\varphi(x,v) - \varphi(y,w) \bigr)\times \\
&\times f(x,v,t)f(y,w,t)C(z) \,dx\,dy\,dz\,dv\, dw \Big \rangle.
 \end{equations}
In \fer{kine-xv} the pairs $(x^*, v^*)$ and  $(y^*, w^*)$ are obtained from the pairs
$(x,v)$ and $(y,w)$  by \fer{k1} and \fer{eq.cpt2}.  Note that, by choosing $\varphi$
independent of $v$, that is $\varphi=\varphi(x)$, equation \fer{kine-xv} reduces to
the equation \fer{kine-w} for the marginal density of conviction $F(x,t)$.

To obtain analytic solutions to the Boltzmann-like equation \fer{kine-xv} is
prohibitive. The main reason is that the unknown density in the kinetic equation
depends on two variables with different laws of interaction. In addition, while the
interaction for conviction does not depend on the opinion variable, the law of
interaction for the opinion does depend on the conviction. Also, at difference with
the one-dimensional models, passage to Fokker-Planck equations (cf. \cite{PaTo14} and
the references therein) does not help in a substantial way. For this reason, we will
resort to numerical investigation of \fer{kine-xv}, to understand the effects of the
introduction of the conviction variable in the distribution of opinions.

\section{Numerical experiments}\label{nume}

This section contains a numerical description of the solutions  to the Boltzmann-type
equation \fer{kine-xv}. For the numerical approximation of the Boltzmann equation we
apply a Monte Carlo method, as described in Chapter 4 of \cite{PT13}. If not otherwise
stated the kinetic simulation has been performed with $N=10^4$ particles.

The numerical experiments will help to clarify the role of conviction in the final
distribution of the opinion density among the agents. The numerical simulations
enhance the fact that the density $f(x,v,t)$ will rapidly converge towards a
stationary distribution \cite{PT13}. As usual in kinetic theory, this stationary
solution will be reached in an exponentially fast time.

The numerical experiments will report the joint density of conviction and opinion in
the agent system. The opinion variable will be reported on the horizontal axis, while the conviction variable will be reported on the vertical one. The color intensity will refer to the concentration of opinions. The following numerical tests have been considered.

\begin{figure}[t]
\includegraphics[scale=.40]{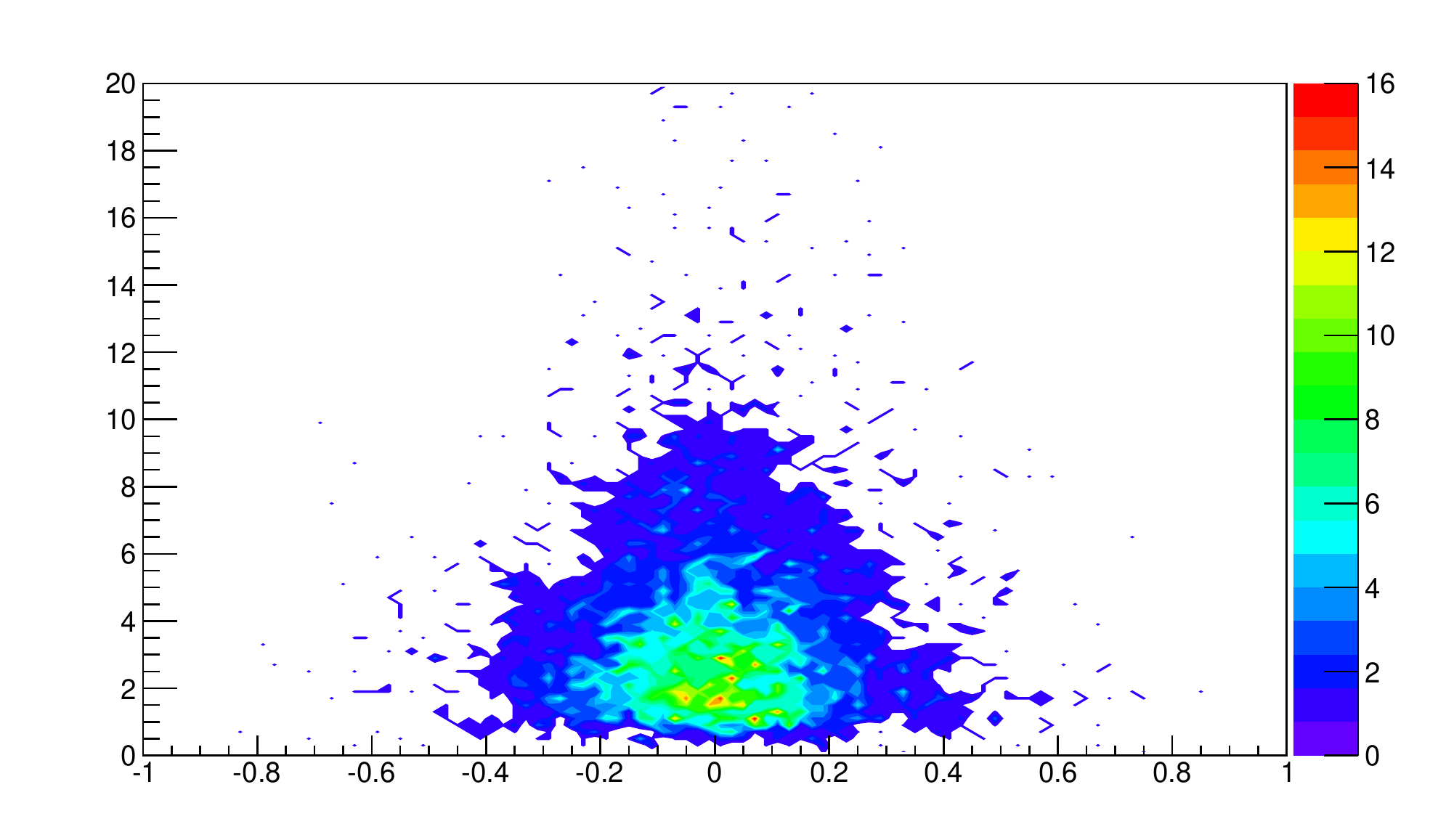}
\includegraphics[scale=.40]{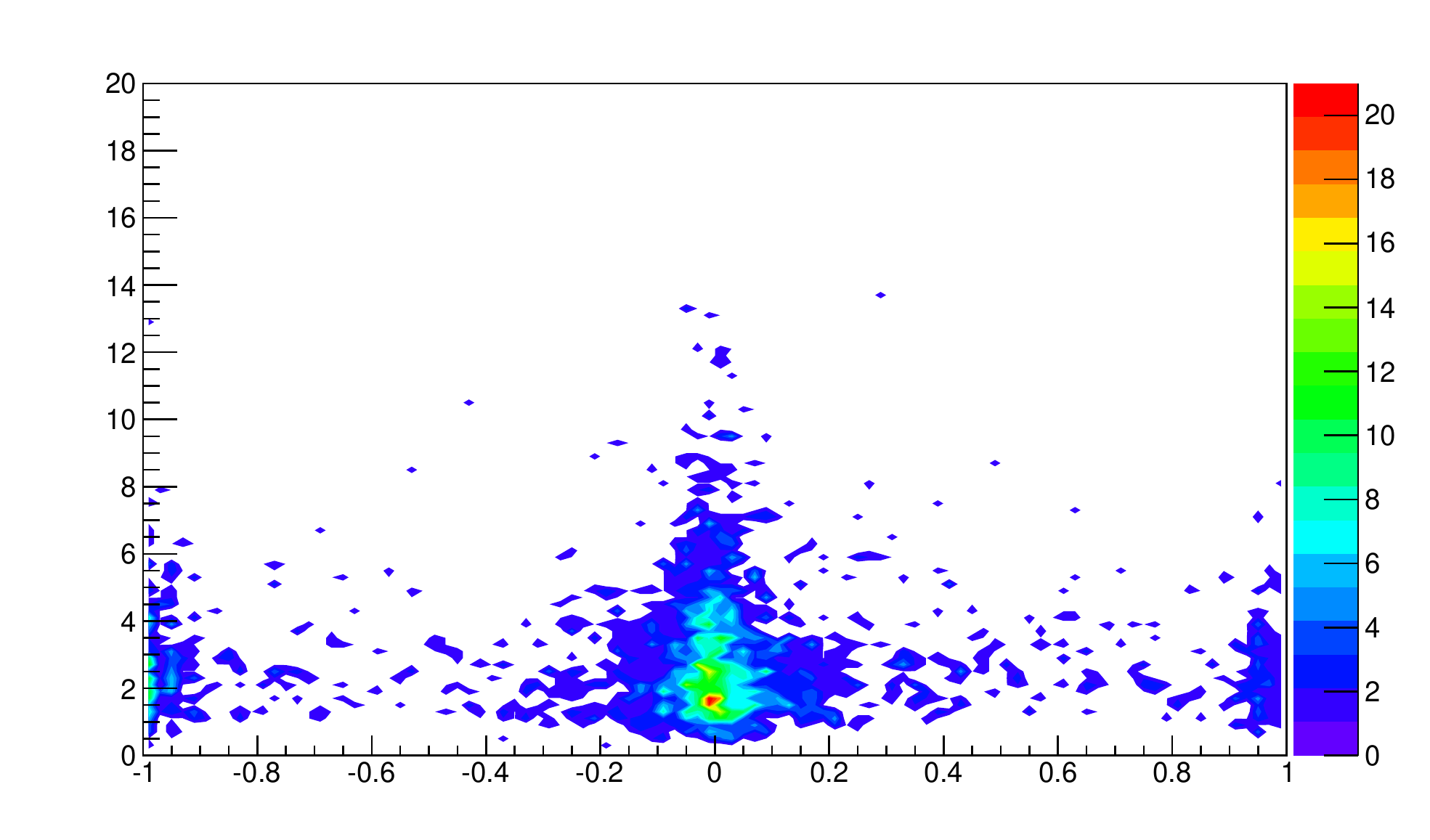}
\caption{Test 1: The particles solution with  $N=10000$ particles and linear $H$. High diffusion in conviction and reduced self-thinking (up) compared to low diffusion in conviction and high self-thinking (down) }
\label{fg:fig1}
\end{figure}
\begin{figure}
\includegraphics[scale=.40]{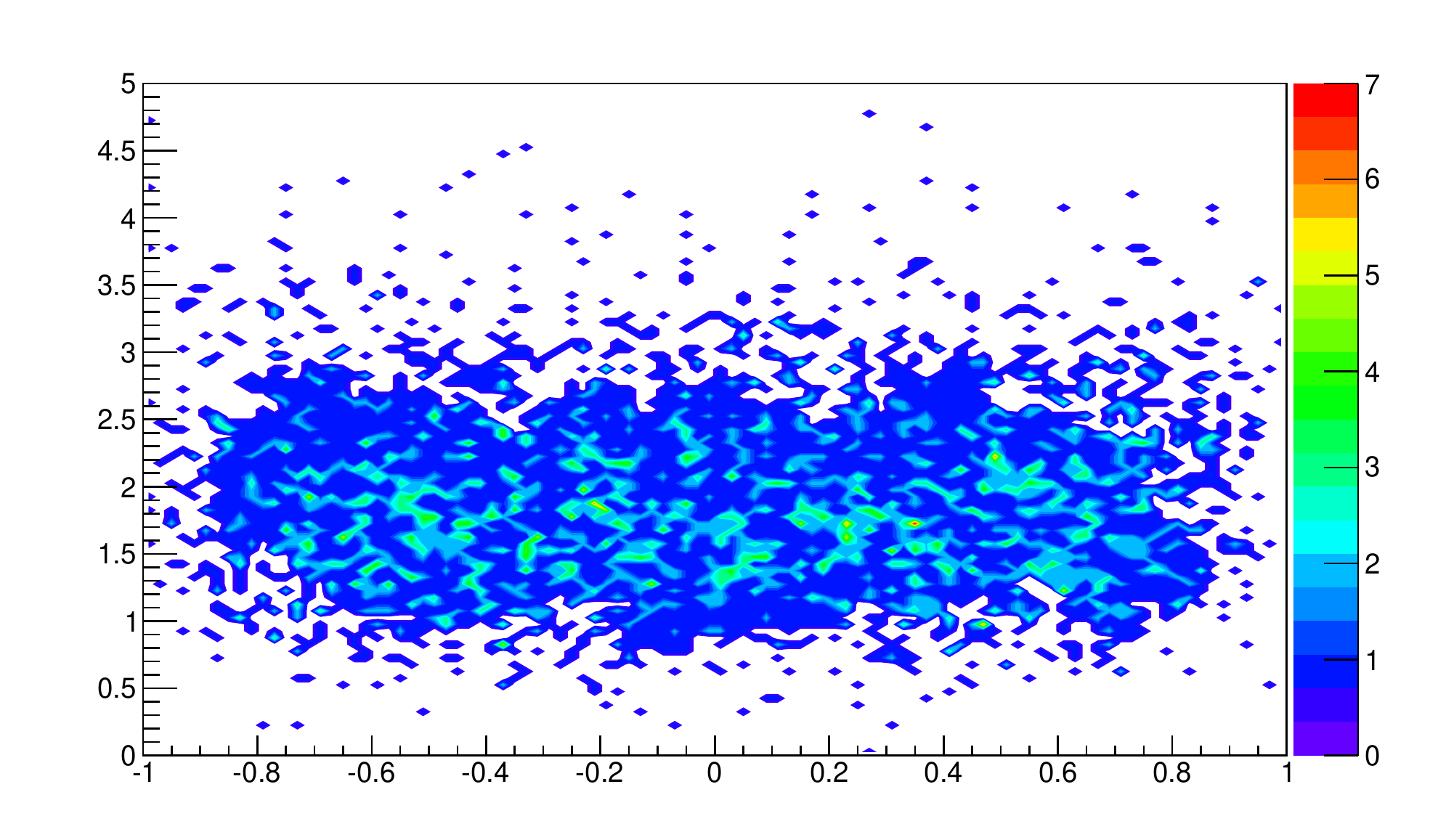}
\includegraphics[scale=.40]{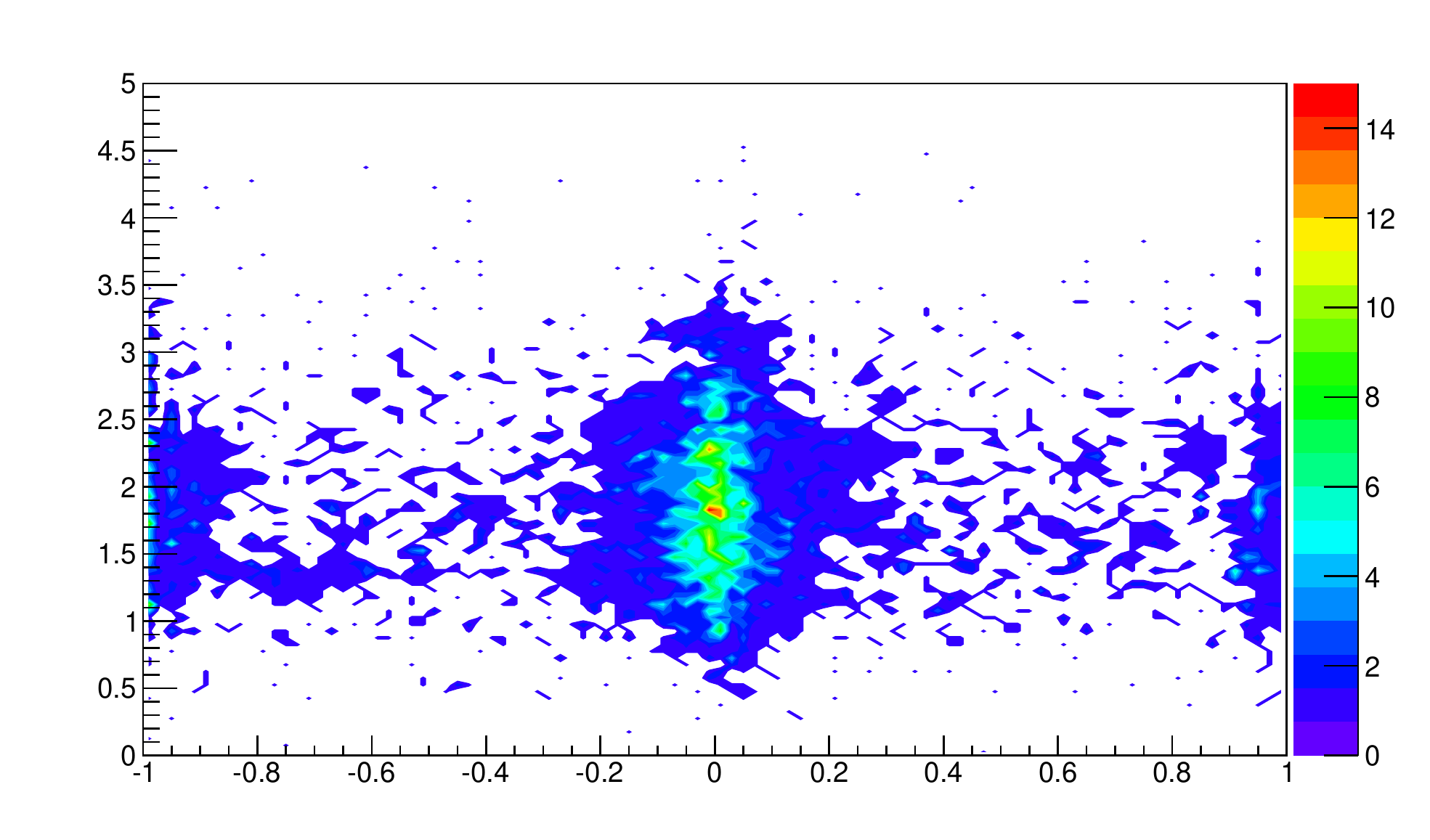}
\caption{Test 2: The particles solution with  $N=10000$ particles and $H(x) = \sqrt{x}$. High diffusion in conviction and reduced self-thinking (up) compared to low diffusion in conviction and high self-thinking (down).}
\label{fg:fig2}
\end{figure}

\subsubsection*{Test 1}
In the first test we consider the case of a conviction interaction where the diffusion
coefficient in \fer{k1} is linear, $H(x)=x$.  As described in Section \ref{know6} the
distribution of conviction in this case is heavy tailed, with an important presence of
agents with high conviction, and a large part of the population with a mean degree of
conviction. In \fer{eq.cpt2} we shall consider
\[
\Phi(x)= \Psi(x) = \frac 1{1+x}.
\]
We further  take $\lambda=\lambda_B=0.5$ in \fer{k1}, and $P(|v|)=1$, $D(|v|)
=\sqrt{1-v^2}$ in \fer{eq.cpt2}.  We consider a population of agents with an initially uniformly distributed opinion and a conviction uniformly distributed on the interval $[0,5]$. We choose a time step of $\Delta t=1$ and a final
computation time of $t=50$, where the steady state is practically reached. 

\begin{figure}[t]
\includegraphics[scale=0.40]{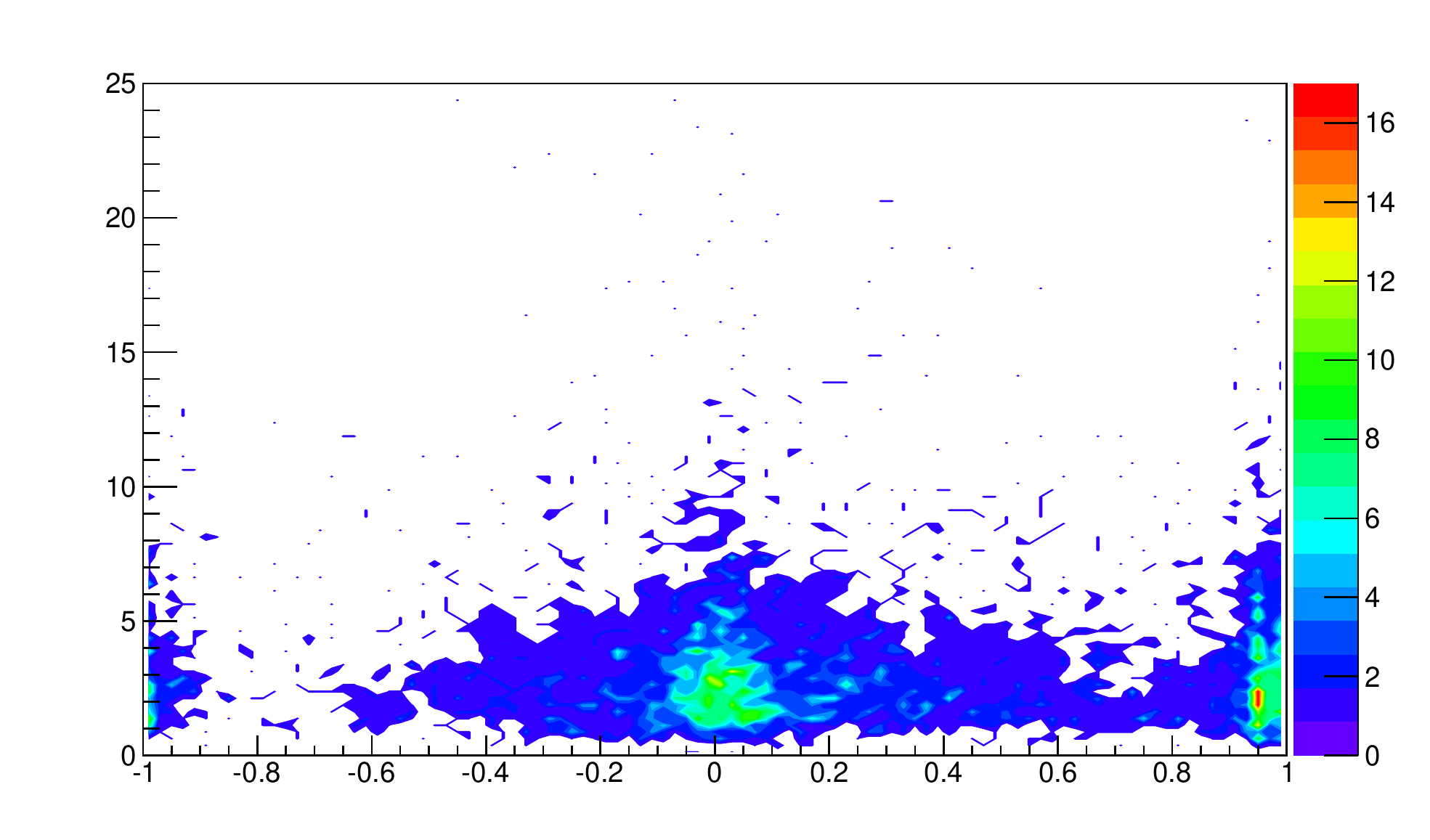}
\includegraphics[scale=0.40]{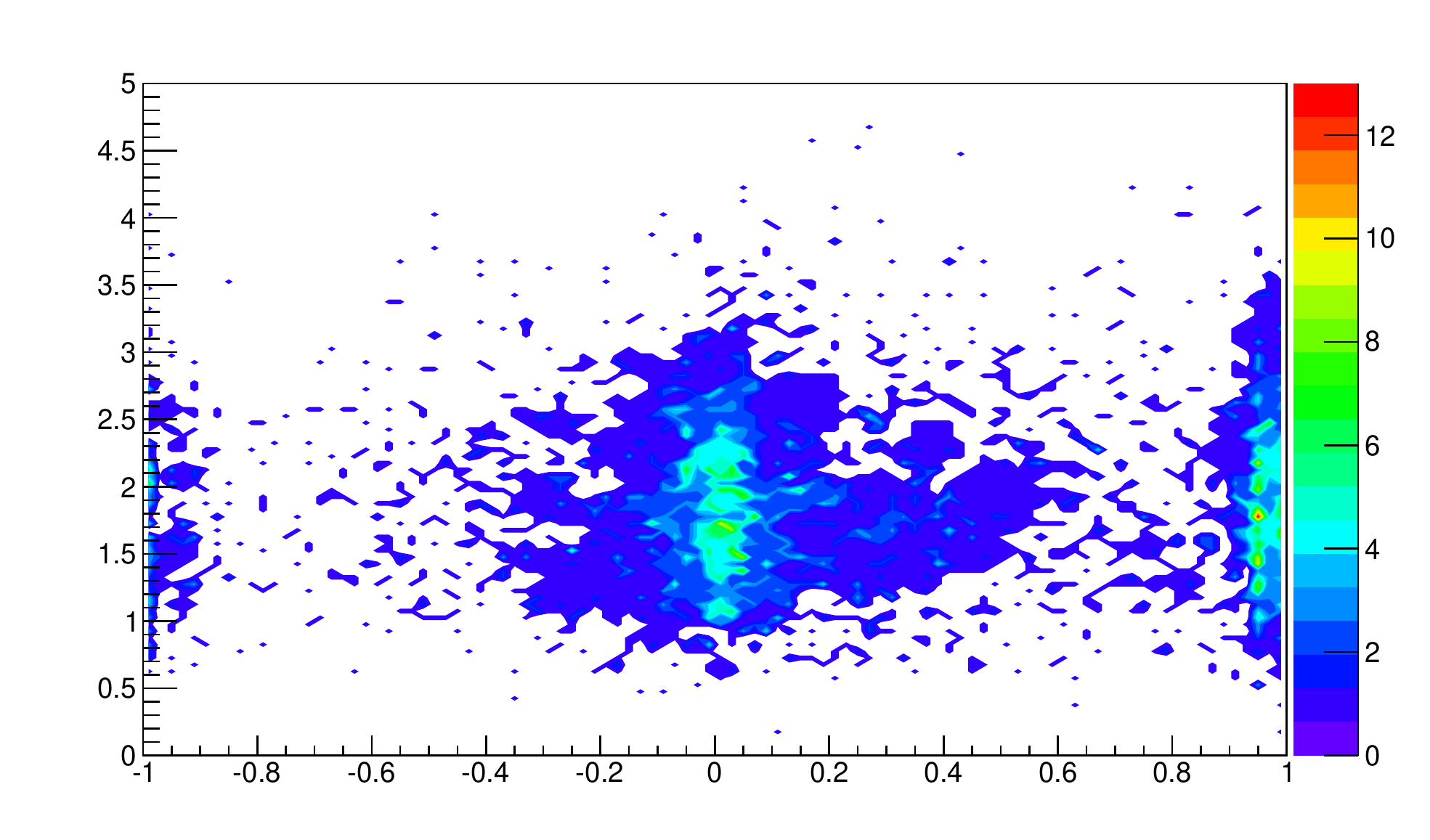}
\caption{Initial asymmetry in opinion leads to different opinion-conviction distributions. $H(x) = x$ (up), and $H(x) = \sqrt{x}$ (down).}
\label{fg:fig3}
\end{figure}

Since the
evolution of the conviction in the model is independent from the opinion, the latter
is scaled in order to fix the mean equal to $0$. We report the results for the
particle density corresponding to different values of $\mu$, $\gamma$ and the
variance $\sigma^2$ of the random variables $\Theta$ and $\tilde\Theta$ in Figure
\ref{fg:fig1}. This allows to verify the essential role of the diffusion processes in conviction and opinion formation.  In Figure \ref{fg:fig3} we plot the marginal
densities together with the tail distribution
\[
{\bar{\mathcal{F}}}(x)=1-\mathcal{F},\qquad {\bar{\mathcal{G}}}(x)=1-\mathcal{G},
\]
which are plotted in \emph{loglog} scale to visualize the tails behavior.

\subsubsection*{Test 2}
In this new test, we maintain the same values for the parameters, and we modify the
diffusion coefficient in \fer{k1}, which is now assumed as $H(x)=\sqrt x$. Within this
choice, with respect to the previous test we expect the formation of a larger class on
undecided agents.  The results are reported in Figure \ref{fg:fig2} for the full
density. At difference with the results of Test 1, opinion is spread out almost uniformly among people with low conviction. It is remarkable that in this second test, as expected, conviction is essentially distributed in the interval $[,5]$, at difference with Test 1, where agents reach a conviction parameter of $20$.

The same effect is evident in Figure \ref{fg:fig3}, which refers to both Tests 1 and 2 in which, to understand the evolution in case of asymmetry, the initial distribution of opinions was chosen  uniformly distributed on the positive part of the interval.

\section{Conclusions}

Opinion formation in a society of agents depends on many aspects, even if it appears
to have very stable features, like formation of clusters.  In this note, we introduced
and discussed a kinetic model for the joint evolution of opinion in presence of
conviction, based on the assumption that conviction is a relevant parameter that can
influence the distribution of opinion by acting on the personal attitude to
compromise, as well as in limiting the self-thinking.  Numerical experiments put in
evidence that the role of conviction  relies in concentrating the final distribution of opinions towards a well-defined one.

\medskip
{\bf Acknowledgement.} This work has been done under the activities of the National
Group of Mathematical Physics (GNFM). The support of the MIUR Research Projects of
National Interest {\em ``Optimal mass transportation, geometrical and functional
inequalities with applications''} is kindly acknowledged. One of the authors (G.T.),
thanks Eitan Tadmor for the interesting discussion about the role of conviction in
opinion formation, which was the starting point for this research.

\end{document}